\documentclass[12pt]{article}
\usepackage{graphicx}
\usepackage[top=3cm, bottom=2cm, left=3cm, right=3cm]{geometry}
\title{Structures in the cosmic ray energy spectra}
\author{$^{1}$ A. D. Erlykin $^{a,b}$ and A. W. Wolfendale $^{b}$\\
$(a)$ P N Lebedev Physical Institute, Moscow, Russia\\
$(b)$ Department of Physics, Durham University, Durham, UK}
\begin{document}
\maketitle

\footnote{Corresponding author: tel +74991358737 \\ 
E-mail address: erlykin@sci.lebedev.ru}

\begin{abstract} 
All the components of Cosmic Rays (CR) have
'structure' in their energy spectra  at some level, ie deviations from a
simple power law, and their examination is relevant to the origin of the
particles. Emphasis, here, is placed on the large-scale structures in the spectra of 
nuclei (the 'knee' at about 3 PeV), that of electrons plus positrons (a shallow 
'upturn' at about 100 GeV) and the positron to electron plus positron ratio (an upturn 
starting at about 5 GeV). 

Fine structure is defined as deviations from the smooth spectra which already allow for
 the large-scale structure. Search for the fine structure has been performed in the 
precise data on positron to electron plus positron ratio measured by the AMS-02 
experiment. Although no fine structure is indicated, it could in fact be present at the
 few percent level. 
\end{abstract}

\maketitle

\section{Introduction}
   Starting with the 'all-particle spectrum' it has been known for many
decades that a simple power law does not represent the spectrum, rather, 
the power law exponent increases at several PeV ( the 'knee') and
falls again at several EeV (the 'ankle'). Both have been examined in
great detail, but without a concensus as to their detailed origin. Here,
as for the all-particle spectrum, we restrict attention to the knee and 
its' apparent 'sharpness'.
   
Turning to electrons, their energy  spectrum has a distinctive shape,
with increasing agreement that, when plotted as $log IE^3$ vs $log E$,
there is a flattening at about 100 GeV and a downturn commencing at
about 1000 GeV. Interest centres on the role of a single source, a
pleasing result in view of the lukewarm reaction to our early 'Single
Source Model' for the knee \cite{EW0}. Surely, the single source responsible for the
flattening in the electron plus positron spectrum has to be different from that 
responsible for the knee, but the principle is the same: a single source is providing 
structure.

 Finally, with respect to positrons, a number of observations have
reported an upturn in the positron fraction (~the number of positrons divided by that 
of  electrons plus positrons~), starting at about 5 GeV
\cite{PAMELA1,ATIC,HESS,FERMI-LAT,AMS-02}. The comprehensive list of references can be 
found in the review \cite{Panov}. The case here for a local source 
providing the extra positrons seems overwhelming.
   In each case, we wish to work out the probability that a single source
(SNR) can produce the observed feature, or one that is at least as
'dramatic'. A small value of the probability will indicate the
unlikelihood of this explanation and a large value, say bigger than a
few tens of percent, will give confidence.
   A related analysis will also be made of techniques for determining to
what extent very precise measurements of apparently smooth energy spectra
can give information about the origin and propagation of the particles.

\section{Large-scale structures in the Cosmic Ray Energy Spectra}
\subsection{The knee in the all-particle spectrum}
   In view of its' history (it was discovered in 1958 \cite{Kulik}) 
this feature is the most studied. In our detailed analysis using the 
developed SNR model \cite{EW1}, a model in which the standard Fermi acceleration 
mechanism was used for acceleration by the SNR shock, we derived the age-distance 
diagram for the SNR
which could be responsible for the formation of the knee, ie our 'single
source' \cite{EW2}. The 95\% confidence level encompassed distance: 250 to 400 pc
and age: 85 to 115 ky. Surprisingly, perhaps, the area occupied by the
age-distance contours for an SNR giving a knee sharper, or more
pronounced, than that observed is not much bigger than this, assuming 
that the energy injected by the SNR into CR is the standard 
$10^{50}$erg. The mean number of SNR expected in the required age-distance region
is about 0.01-0.02, ie the probability to find SNR within this region, which gives the 
sharp knee-like structure, is about 2\%.

If we relax the requirement of the sharpness and look for somewhat smoother 
deviations from the simple power law then the examination of simulated
spectra presented in \cite{EW3} (~anomalous diffusion with $\alpha=1$~) gives the 
probability of such features of about 28\%. In view of the difficulty in both measuring
 the knee and in interpreting it (~diffusion characteristics, etc~) we see no reason to
 doubt the conclusion that a single source is responsible. 

\subsection{The electron component}
   In \cite{EW4} we presented a model in which, as for protons, electrons are 
accelerated by the shock in the SNR, the SNR then being distributed in the Galaxy 
randomly in space and time. The well-known steeper energy spectrum for electrons than 
for protons was explained by way of an energy-dependent Mach number for the shock. It 
must be remarked, however, that this feature is not yet fully understood. 

Inspection of our model's prediction for the shape of the electron
energy spectrum \cite{EW4} shows a wide range of spectra with a median
intensity ($logIE^3$) which falls slowly with energy to about 100 GeV,
beyond which it falls rapidly. This is in contrast with the observed
electron plus positron spectrum which rises slightly or flattens at about 100GeV and
falls with modest rapidity above 1000 GeV. It should be noted that the
ATIC spectrum \cite{ATIC} has remarkable structure in the range 100 to 1000 GeV,
but this is not shared by other measurements, eg Fermi LAT \cite{FERMI-LAT}.
  Examination of \cite{EW4} indicates that about 24\% of the predicted spectra
have the necessary or stronger large-scale spectral structures.
\subsection{The positron fraction}
   A number of experiments have shown an upturn in the positron fraction, as
mentioned already. The most precise are due to : PAMELA \cite{PAMELA1}, Fermi-LAT
\cite{FERMI-LAT} and AMS-02 \cite{AMS-02}. Many authors have suggested that a pulsar is
responsible , but, in a very recent work \cite{EW5} , we have proposed the SNR
presumed to be the predecessor of the pulsar Geminga. In this case, the positrons come 
from radioactive ejecta from the SN; the positrons are then accelerated by the SNR 
shock in the usual way. An advantage of this mechanism is that the acceleration 
efficiency of the near-1 MeV positrons is very high.

   In a manner similar to that for the origin of the knee (~\S2.1~),
we have determined the limits on distance $D$ and age $T$: $250 < D < 320$ pc and 
$170 < T < 380$kyear. The mean number of SN explosions
expected in this age-distance range is about 0.3 and the probability of just a single
 occurence according to the Poisson distribution is about 22\% which is not an
unreasonable value. Eventual measurements of the anisotropy of electron and positron 
arrival directions will show whether, or not, the identification is correct.

\subsection{Anti-protons}
   If anti-protons are generated and accelerated in SNR, by the
interactions of protons and heavier nuclei with the interstellar medium (ISM) within 
the remnant, then their spectrum should show large-scale structure, in the form of an 
upturn in the \={p}/p ratio. Interestingly, neutrons produced in \={n},n - pairs
in the interactions will augment the \={p} flux. The neutrons (and \={n})
have the advantage of escaping the magnetic trapping, which may be strong
in the early remnant, before decaying into p and \={p}.

The only measurements extending as far as hundreds of GeV
are those from PAMELA \cite{PAMELA3}, which finish at 180 GeV, although the errors
are large at the high energies. The measured ratio at 100 GeV is 
$(1.7\pm 0.5) \cdot 10^{-4}$, but this includes a rather uncertain background
contribution \cite{PAMELA3} so that a single source contribution, or upturn in
the ratio, cannot yet be determined. All that can be said at present is
that the \={p}/p ratio hints at a flattening above 10 GeV, which, if confirmed and 
after subtraction of the rapidly falling background and accounting for a steep proton 
spectrum in the denominator of \={p}/p ratio, would suggest the onset of a finite 
single source contribution. 

Its absence would suggest that \={p} are not produced by the local SNR, unlike the 
positrons which we hypothesise to come from the radioactive SN ejecta. At present the 
expected 'cross-over energy', where SNR-generated \={p} equal background, is not clear.

\section{Fine Structure in the Positron Fraction}
\subsection{Search for fine structure}

    By 'fine' rather than 'large-scale' structure we mean anything in the spectrum, or 
particle (positron) fraction, that is over and above the first order fit, suggested by 
the AMS-02 collaboration \cite{AMS-02}. This fit  
was the sum of two simple expressions: a power law for those positrons 
coming largely from CR interactions in the ISM and an exponentially 
modified power law for positrons from a local source. The 
energy range is from 1 to 300 GeV, ie $logE,GeV$: 0 to 2.5.
     Below 10 GeV , solar modulation is important  and thus we divide 
the data into two parts: $logE,GeV$ =0 to 1.25 and $logE,GeV$ = 1.25 to 2.5.
The lower part might show fine structure due largely to solar modulation, whereas 
the higher energy part might indicate Galactic effects, associated with 
the finite number of sources contributing to the CR flux.

In an attempt to determine at least an upper limit to the fine structure in 
the positron fraction we have examined the 'precise' AMS-02 data in some 
detail, as follows.
     Fits to the ratio of the measured positron fraction to that obtained by its fit
with the AMS-02 suggested function were made for the two halves of the data for 
various degrees of polynomial function 'n' from 2 to 9 with particular emphasis on 2 
and 9 themselves. Thus, we are searching for fine structure within the already 
allowed - for large-scale structure. Clearly, it will be small, otherwise it 
would have been commented on already.
\begin{figure}[htb]
\begin{center}
\includegraphics[height=14cm,width=7cm,angle=-90]{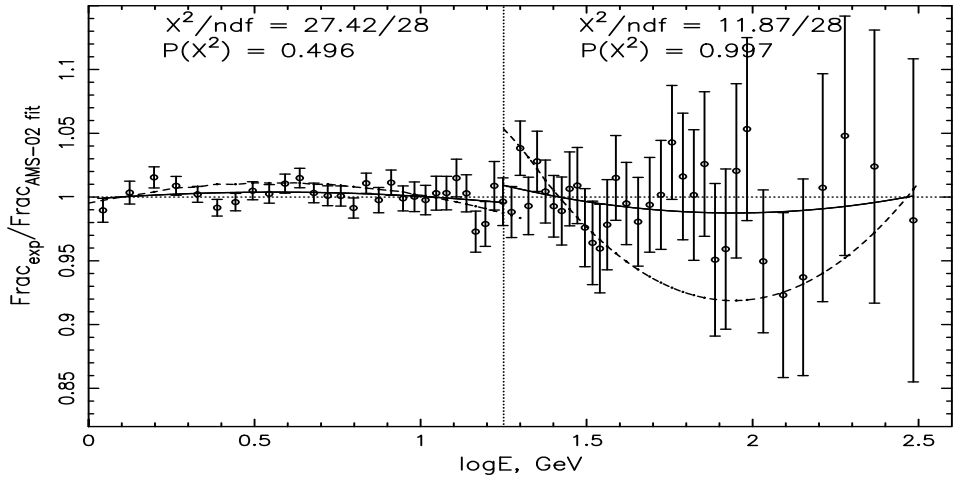}
\end{center}
\caption{\footnotesize The ratio of the positron fraction measured in the AMS-02 
experiment to its fit by the AMS-02 suggested function. Errors of the ratio are 
statistical. Full lines in both halves of the energy range show the weighted fit of the
 ratio by the 2-degree polynomial function: $a_0 + a_1X +a_2X^2$, where $X = logE$. The
 quality of the fit is shown by the values of the reduced $\chi^2$. Dashed lines 
indicate the structure, which is of the same shape of the 2-degree polynomial, but has 
 $P(\chi^2) = 0.05$.}   
\label{fig:fig1}
\end{figure}

     Figure 1 shows the results for n = 2 (full lines) and the derived chi-square 
value with its' significance for both halves of the studied energy range. 
Indicated errors of the ratio are statistical. Systematic errors are weakly energy 
dependent and cannot have irregular behaviour with the energy. In this analysis we did
not take them into account since they cannot help to reveal the possible fine structure
of the energy spectra. 

The weighted polynomial best-fit has a reasonable significance at lower energies and 
the very large value of the probability at high energies indicates that even the
statistical errors shown may have been overestimated. The limit for a 5\% probability 
of fit is shown by the dashed 
lines. The 5\% level is the usual value for 
acceptance of a fit as being just non-allowable. It will be noted that 
the 'fine structure' of this shape could reach about 8\% maximum at least in the high 
energy part of the range and still not be dis-allowed by the data. We presented these 
dashed lines just as examples illustrating the non-negligible probability of  
deviations from the AMS-02 fit at the level of the few percent.
 
    Figure 2 shows the situation for n = 9. The best-fit has again a 
very high probability in both halves of the range. This fit has some 
interesting features although none is as yet significant. The 5\% probability limits 
are shown by the dashed lines. They show that the amplitude of 'undulations' in some 
restricted regions are still allowed to be as high as 17\% by the data. Again the 
dashed lines here are given as possible examples of the fit which still have the 
allowable probability.
\begin{figure}[htb]
\begin{center}
\includegraphics[height=14cm,width=7cm,angle=-90]{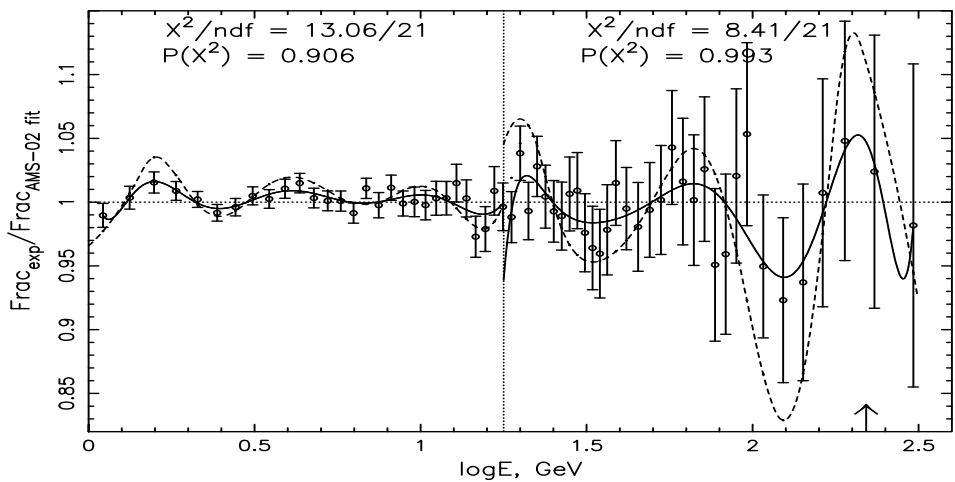}
\end{center}
\caption{\footnotesize The same as in Figure 1, but the ratio is fitted by the 
9-degree polynomial function. The position of the minimum in the ATIC electron plus 
positron energy 
spectrum \cite{ATIC} is indicated by the arrow. Its approximate coincidence with the 
upward excursion of the positron fraction from the regular model calculation is an 
interesting feature. However, due to the low statistics, it should be regarded rather
as a hint for the possible fine structure.}   
\label{fig:fig2}
\end{figure}

\subsection{Discussion of possible sources of fine structure}
\subsubsection{Solar wind modulations}

  As remarked already, the lower energy region is the province of solar 
modulation, which is known to be dependent on the charge sign of the CR and on 
the time of observation, by way of the solar magnetic field polarity 
[eg \cite{Della}]. The difference between the positron fraction measured by AMS-01
\cite{AMS-01} and PAMELA \cite{PAMELA1} in the different periods of solar activity
can be caused by this phenomenon \cite{Panov}. 

Although precise measurements of AMS-02 in the lower
energy region are quite consistent with their fit by two-term function (~background 
plus single source, $\chi^2/ndf = 28.66/31, P(\chi^2) = 0.586$~) it is worth-while to 
estimate the upper limits of the possible fine structure consistent with these 
measurements. As can be seen in Figure 1 the maximum contribution of the 2-degree
polynomial-like structure which still has the 5\% confidence level of consistency with 
data points is about 0.011. The same estimate for the narrower structures which appear 
in the 9-degree polynomial fit gives a maximum contribution of only 0.035. 

It is seen that the present high precision measurements does not allow the presence 
of fine structure greater than about 3-4\% in the GeV energy region due to solar 
wind modulation. Later 
measurements should be of adequate accuracy to enable the effect of time-dependent 
modulation to be studied, using factors such as those in \cite{Katsa}.

\subsubsection{Galactic source modulations}
  We now discuss the higher energy region.
  It is here that the ATIC results \cite{ATIC} have relevance, in that the 
measured electron plus positron spectrum has considerable structure. The energy of 
about 220 GeV at 
which an apparent minimum in the electron + positron intensities appeared is 
indicated in Figure 2. It would have been expected that, if positrons were 
uncorrelated, there would have been a maximum at this energy in the positron fraction. 
However, statistical errors of the measured positron fraction are too high to confirm 
the anti-correlation between the ATIC minimum of the electron plus positron intensity 
and the AMS-02 maximum of the positron fraction. Other ATIC minima are at higher 
energies which do not overlap with the range of AMS-02 measurements.  

\subsubsection{Spectral structure as a diagnostic of CR origin}
Both SNR and pulsars are candidates for CR origin and a distinction between them is not
 a trivial problem. However, spectral structure (~or fine structure~) can be useful, 
particularly for the electron component which, because of energy losses, comes 
predominantly from 'local' regions (~within a kpc or so~) and thus from a smaller 
number of CR sources.

A comparison can be made between our electron spectra from the random SNR model 
\cite{EW4} and the prediction for pulsars \cite{Panov,Malysh}. It is immediately 
apparent that the prediction for SNR are 'smoother' than those for pulsars. Over the 
range 100-2000 GeV, the SNR model has rarely excursions in 'intensity' ($logIE^3$) 
bigger than 30\% whereas for pulsars there are four excursions with a mean of 40\%.

The reason for the difference is self-evident. For SNR, in the model, at least, unique 
energy spectra are emitted from the SNR when the SNR 'bubble' bursts, and the spectral
structure arises from propagation effects alone, ie contributions from SNR of different
ages at different distances. For pulsars, the flatter 'emission spectra' have maximum 
energies depending on their ages, with, conventionally, sharp cut-offs, and these 
fluctuations are added to those due to propagation.

Comparing the structure for electrons and protons, it is useful to examine the range of
 intensities as a function of energy (~for the same propagation model~) from our work
\cite{EW3,EW4}, which relates to 50 independent samples. A large range suggests more
structure than is the case for a small range. It is found that the ranges for electrons
 and protons are similar to about 1000 GeV, above which the range for electrons 
increases more rapidly. A similar result is apparent for the degree of 'oscillation' 
of the spectra - that for electrons is singularly large in the next decade of energy. 
However, the 'degree' of oscillation is hard to quantify and this is why the range of 
intensities is considered. This behaviour follows from the fact that electron losses 
increase as the energy squared and the transit time from source to observer; the actual
spatial and temporal distribution of nearby sources is therefore critical.
\section{Conclusion}
We have analysed the available results on the energy spectra of CR particles from the 
standpoint of the 'structures' in their energy spectra. Large-scale structures are 
regarded as differences from simple forms, which point to the existence of a single 
SNR. The model adopted is that introduced by us \cite{EW3,EW4} 
involving CR origin from randomly situated SNR from which CR diffuse. The probability 
of seeing such structure is $\sim30$\% for nuclei, $\sim 20$\% for electrons and 
positrons. For anti-protons, measurements cease at the energy at which structure might 
be expected to show itself.; more extended measurements might show an upturn in the 
\={p}/p ratio.

Fine structure is defined as deviations from the smooth spectra which already allow for
 the large-scale (single source) structure. The precise positron fraction data 
\cite{AMS-02} are 
taken as an example. The datum is taken as the two components: background plus single 
source. The polynomial fits are taken as examples: n = 2 and n = 9, and the data are 
divided  into equal energy ranges: $logE, GeV = 0 - 1.25$ and $1.25 - 2.50$. Although 
no fine structure is indicated , it could be present at the few percent level. For the 
lower energy band solar modulation effects, which are charge dependent, should be 
detectable when temporal and somewhat better statistical data are available. 

For the higher energy range, models are not yet available for the fine structure 
expected as a result of detailed source mechanisms (eg SNR or pulsars) and 
irregularities of CR diffusion, but these will come. Again, although the positron 
fraction data are statistically precise, fine structure could be present at a few
percent level (~up to 8\% for n = 2 and up to 17\% for n = 9 as an example~).

\vspace{5mm}

{\bf Acknowledgements}

The authors are greatful to the Kohn Foundation for financial support.


\begin{thebibliography}{99}
\bibitem{EW0}    Erlykin A.D. and Wolfendale A.W., J. Phys. G: Nucl. Part. Phys. 23 
(1997) 979
\bibitem{PAMELA1} Adriani O. et al., 2009, Nature (London), 458, 607; arXiv:1103.2880
\bibitem{ATIC} Panov A.D. et al., 2011, Astrophys.Space Sci. Trans. 7, 119
\bibitem{HESS} Aharonian F. et al., 2008, Phys. Rev. Lett. 101: 261104; arXiv: 
0811.3894
\bibitem{FERMI-LAT} Ackermann M. et al., 2010, Phys. Rev. D 82, 092004
\bibitem{AMS-02} Aguilar-Benitez M. et al., 2013, Phys. Rev. Lett., 110, 141102.
\bibitem{Panov} Panov A.D., 2013, J. Phys. Conf. Ser. 409, 012004; arxiv:1303.6118
\bibitem{Kulik} Kulikov G.V. and Khristiansen G.B. 1958, J. Exp. Theor. Phys. 35, 635
\bibitem{EW1} Erlykin A.D. and Wolfendale A.W., 2001, J. Phys. G: Nucl. Part. Phys. 27,
 941
\bibitem{EW2} Erlykin A.D. and Wolfendale A.W., 2003, J. Phys. G: Nucl. Part. Phys. 29,
 709
\bibitem{EW3} Erlykin A.D. et al., 2003, Astropart. Phys. 19, 351
\bibitem{EW4} Erlykin A.D. and Wolfendale A.W., 2002, J. Phys. G: Nucl. Part. Phys. 28,
 359
\bibitem{EW5} Erlykin A.D. and Wolfendale A.W., 2013, Astropart. Phys., 49, 23; doi:
10.1016/j.astropartphys.2013.08.001; arxiv:1308.4878
\bibitem{PAMELA3} Adriani O. et al., 2013, J. Exp. Theor. Phys. Lett., 96, 621
\bibitem{Della}  Della Torre S. et al., 2012, Adv. Space Res. 41, 1587
\bibitem{AMS-01} Aguilar M. et al., 2007, Phys. Lett. B 646, 145
\bibitem{Katsa} Katsavrias C. et al., 2012, Adv. in Solar Phys., 280, 623;   
doi:10.1007/s11207-012-0078-6
\bibitem{Malysh} Malyshev D. et al., 2009, Phys. Rev. D 80, 063005 
\end{thebibliography}
\end{document}